\documentclass{ws-procs10x7}
\usepackage{balance}

\newcolumntype{d}[1]{D{.}{.}{#1}}

\def\Journal#1#2#3#4{{\it #1} {\bf #2}, #3 (#4)}

\makeindex
\begin{document}

\title{NEW RELATIONS BETWEEN DIS SUM RULES}

\author{A. L. KATAEV$^*$}

\address{Institute for Nuclear Research 
of the Academy of Sciences of Russia,
Moscow, 117312, Russia\\$^*$E-mail: kataev@ms2.inr.ac.ru}

\twocolumn[\maketitle\abstract{
New relations between Bjorken polarized,
Gross-Llewellyn Smith and   Bjorken unpolarized 
sum rules 
are proposed. They are based on the  ``universality'' of the 
  perturbative  and 
non-perturbative  $\rm{1/Q^2}$ contributions to these sum rules.
The letter facts  can be deduced from the corresponding   
renormalon calculations. The similarity of $\rm{1/Q^2}$ corrections 
are checked by inspecting the numerical  results  
obtained  within several approaches.
The discussed   relations 
are in agreement with 
existing experimental data.Some possible new phenomenological 
applications are mentioned including estimates of not yet measured 
Bjorken unpolarized sum rule.}
]   

There are   approximate relations between the     
Bjorken sum rule for polarized charged leptons-nucleon DIS
($\rm{Bp}$), the 
Gross-Llewellyn Smith sum rule ($\rm{GLS}$) and  
Bjorken unpolarized sum rule ($\rm{Bup}$) for $\nu N$
DIS ~\cite{Kataev:2005ci},~\cite{Kataev:2005hv}, namely 
\begin{equation}
\rm{Bp}(Q^2)\approx \frac{g_A}{18}\rm{GLS}(Q^2) 
\approx \frac
{g_A}{6}\rm{Bup}(Q^2)
\label{rel}
\end{equation}
They  include   perturbative 
QCD corrections and  non-perturbative $\rm{1/Q^2}$-effects 
and  are valid in the case when  $\rm O(1/Q^4)$ terms 
can be neglected. 
Eq.(\ref{rel}) was discovered by  analysing  the  
renormalon model calculations of Refs.~\refcite{Broadhurst:1993ru},
~\refcite{Broadhurst:2002bi}.
The definitions of the sum rules we are interested in 
are well-known: 
\begin{eqnarray} 
\nonumber
{\rm GLS}&=&\frac{1}{2}\int_0^1  \bigg[\rm{F_3^{\nu n}}(\rm{x,Q^2})+
\rm{F_3^{\nu p}}(\rm{x,Q^2})\bigg]\rm{dx}~~, \\ \nonumber  
{\rm Bp}& =&\int_0^1 \bigg[\rm{g_1^{lp}}(\rm{x,Q^2})-\rm{g_1^{ln}}(\rm{x,Q^2})
\bigg]\rm{dx}~~, 
\\ \nonumber 
{\rm Bup}&=&\int_0^1 \bigg[\rm{F_1^{\nu p}}(\rm{x,Q^2})-\rm{F_1^{\nu n}}
(\rm{x,Q^2})\bigg]\rm{dx}~~.
\end{eqnarray}
Within QCD they can be  expressed as  
\begin{eqnarray}
\nonumber
{\rm GLS(Q^2)}&=&3
\bigg[{\rm C_{GLS}}(a_s)-\frac{\rm{A}}
{\rm{Q^2}}
+{\rm O(1/Q^4)}\bigg]  \\ \nonumber
{\rm Bp(Q^2)}&=& \frac{\rm{g_A}}{6}\bigg[{\rm C_{Bp}}(a_s)-\frac{\rm{B}}
{\rm{Q^2}}+{\rm O(1/Q^4)}\bigg] 
\\ \nonumber 
{\rm Bup(Q^2)}&=&{\rm C_{Bup}}(a_s)-\frac{\rm{C}}{\rm{Q^2}}
+{\rm O(1/Q^4)}
\end{eqnarray}   
where $a_s$=$a_s(\rm{Q^2})$=$\alpha_s(\rm{Q^2})/4\pi$ and the  coefficient 
functions 
\begin{eqnarray}
\nonumber 
{\rm C_{GLS}}(a_s)&=&1-4a_s-O(a_s^2) \\ \nonumber 
{\rm C_{Bp}}(a_s)&=& 1-4a_s-O(a_s^2) \\ \nonumber 
{\rm C_{Bup}}(a_s)&=&1-\frac{8}{3}a_s-O(a_s^2)
\end{eqnarray}
are explicitly calculated up to $a_s^3$-corrections 
(for a review see ~\cite{Hinchliffe:1996hc}).
The  non-perturbative parameters $\rm{A}$ 
and $\rm{C}$ of the $\rm{1/Q^2}$ corrections to  the $\nu N$ DIS sum rules  
are connected  to  matrix elements of  operators
composed from quark and gluon fields, and  written down in 
Ref.~\refcite{Shuryak:1981kj}. The numerator $\rm{B}$ of the ${\rm1/Q^2}$ 
term in the  $\rm{Bp}$ sum rule is defined by matrix elements
calculated  in Ref.\refcite{Shuryak:1981pi}, with  additional  
input from Ref.\refcite{Ji:1993sv}. The discussions presented 
below are mainly based on the work
\cite{Kataev:2005hv}. 

The ``universality'' of perturbative contributions to 
the sum rules means  that  the asymptotic structures 
of the expansion of their   coefficient functions in the QCD coupling constant 
is almost identical, namely 
\begin{equation}
\label{pertuv}
{\rm C_{GLS}}(a_s)\approx {\rm C_{Bp}}(a_s) \approx 
{\rm C_{Bup}}(a_s)~~~. 
\end{equation}
The ``universality'' of the  non-perturbative  $\rm{1/Q^2}$-contributions 
to the same sum rules implies that 
\begin{equation}
\label{nonpert}
\rm{A} \approx \rm{B} \approx \rm{C}
\end{equation}
Following \cite{Kataev:2005hv}, consider now  
the method of renormalon calculus,
advanced  in  Ref. \cite{Zakharov:1992bx}, 
and  reviewed in detail in  
~\cite{Beneke:1998ui} and ~\cite{Beneke:2000kc}.  
The  coefficient function of  
DIS sum rules can be presented in the  form of the Borel integral 
\begin{equation}
{\rm C}(a_s)=\int_0^{\infty} {\rm exp}(-\delta/\beta_0 a_s)~
\rm{B}[\rm{C}(\delta)]~{\rm d\delta}
\end{equation}   
where $\rm{B}[\rm{C}(\delta)]$ is the  
Borel transform,
defined as $\rm{B}[\rm{C}(\delta)]$=
$\sum_{n=0}^{\infty}\rm{d_n}(\delta^n/n!)$.  Here  $\rm{d}_n$ are 
the coefficients of the asymptotic perturbative expansion of  
 ${\rm C}(a_s)$  and 
$\beta_0=(11/3)C_A-(4/3)T_fN_f$ is the first coefficient of the 
QCD  $\beta$-function, $C_A=3$, 
$T_f=1/2$ and $N_f$ - number of the active  quarks flavours. 
The usual prescription of calculating the  Borel transform  in QCD is to 
evaluate the contributions of Feynman diagrams with   one gluon line,
dressed by a chain  
of fermion loops, each proportional to  $N_f$. 
Their  contributions 
to the coefficient functions 
do not reflect the whole 
picture of  renormalon effects in QCD. The latter begin to 
manifest themselves after the replacement $N_f\rightarrow -(3/2)\beta_0$=
$N_f-(33/2)$.
For the 
$\rm{GLS}$ and ${\rm Bp}$ sum rules 
the corresponding Borel transforms
coincide~\cite{Broadhurst:1993ru}:
\begin{eqnarray}
\nonumber
\rm{B}[\rm{C_{Bp}}]&=&\rm{B}[\rm{C_{GLS}}] \\  
&=&
-\frac{(3+\delta){\rm exp}(5\delta/3)}{(1-\delta^2)(1-\delta^2/4)}
\end{eqnarray}
The 
Borel transforms for  the ${\rm Bup}$ and ${\rm Bp}$ turn out to be closely 
related~\refcite{Broadhurst:2002bi},   namely  
\begin{eqnarray}
\nonumber
\rm{B}[\rm{C_{Bup}}]&=&\bigg(\frac{2(1+\delta)}{3+\delta}\bigg)
\rm{B}[\rm{C_{Bp}}] \\  
&=& -\frac{2{\rm exp}(5\delta/3)}{(1-\delta)(1-\delta^2/4)}
\end{eqnarray}
Notice two similar features of Eq.(5) and Eq.(6): the 
leading $\delta=1$ poles  in the Borel transforms of Eq.(5) and Eq.(6), 
which are lying on the positive axis of the 
$\delta$-plane and are  called leading infrared renormalons (IRRs),     
have  identical  negative residues. Moreover, the residues of  
$\delta=-1$ poles in the same Borel transforms,
called leading  ultraviolet renormalons (UVRs), are strongly suppressed 
in relation to the residues of the leading IRRs~\cite{Broadhurst:2002bi}. 
Indeed, in the case of Eq.(5) the suppression factor is 
$(1/2){\rm exp}(-10/3)\approx 0.018$, while 
in the case of Eq.(6) it is {\bf identically} equals to {\bf zero}. 

Eq.(5) 
indicates the validity of the l.h.s. of the 
perturbative Eq.(2) 
${\rm C_{GLS}}(a_s)\approx {\rm C_{Bp}}(a_s)$ in the asymptotic regime 
(for  technical details see  
Ref.~\refcite{Broadhurst:1993ru}).
This  suggests that the 
  ``light-by-light''-type  
contributions to high-order perturbative 
corrections   to  ${\rm C_{GLS}}(a_s)$ should be small.
The existing order $a_s^3$ analytical results ~\cite{Larin:1991tj}
confirm this guess and are  underly  definite 
next-to-next-to-leading order  phenomenological applications of 
 Ref.~\refcite{Brodsky:1995tb}. 

The consequence of Eq.~(6) is even more clear. In this case  
the behaviour of  asymptotic series for the coefficient function is 
governed by the same IRR, which dominates the asymptotic perturbative 
expressions
for the coefficient function of the ${\rm Bp}$ and  ${\rm GLS}$
sum rules. Therefore the  r.h.s. of Eq.(2), namely 
${\rm C_{Bp}}(a_s)\approx {\rm C_{Bup}}(a_s)$ is valid as well.
Moreover, this explains the similarity between 
next-to-next-to-leading order perturbative QCD contributions
to  ${\rm C_{Bp}}(a_s)$ and  ${\rm C_{Bup}}(a_s)$ observed 
in Ref.\cite{Gardi:1998rf}. 

A few  words about non-perturbative $1/Q^2$ effects are in order.
The  IRRs  in the Borel transforms  for 
all three  sum rules at $\delta=1$ and 
$\delta=2$  generate ambiguities in the corresponding Borel  
integrals of Eq.(4). 
In our analysis we will  
modify the integration contour by  introducing small 
semi-circles, which are  
going above  these poles in the  Borel plane.
This {\rm PV} procedure  introduse  an  extra  negative 
IRR -induced contribution,
namely 
\begin{equation}
\label{rencont}
\Delta{\rm C}_{sum~rules}\approx -\frac{32{\rm exp}(5/3)}{3\beta_0}
\frac{\Lambda_{\overline{\rm MS}}^2}{\rm{Q^2}}~~~. 
\end{equation}
A similar term was derived in Ref.\refcite{Beneke:2000kc}
in the  context of discussions of the  ${\rm GLS}$ sum rule. 
This IRR induced power correction should be cancelled by 
the leading  UVR  in the coefficient function 
of the twist-4 $1/Q^2$-term, as was mentioned in the work of 
Ref.\refcite{Mueller:1993pa} and explicitly shown in the   
theoretical analysis of   
Ref. \refcite{Gardi:2002xm}.
However, the negative sign and the  identical value of Eq.(7) 
in the case of all three sum rules  
can  be considered as an argument in favour of negative  
and identical  values 
 of the  numerical results for  $1/{\rm Q^2}$ contributions to 
different sum rules.
This statement  
is similar to the hypothesis of universality of  power corrections 
and IRR contributions \cite{Dokshitzer:1995qm} and supports the validity 
of Eq.(3). In   Table 1   
the numerical expressions for the numerators 
of ${\rm 1/Q^2}$-corrections 
calculated   within different 
non-perturbative models are summarized. 

\begin{table}
\tbl{Coefficients of 
the   twist-4 contributions to DIS sum rules (in ${\rm GeV^2}$) \label{tab1}}
{\begin{tabular}{@{}lccr@{}}
\toprule
~~~~~~{\bf A}  & {\bf B} & {\bf C}  
& Ref.  \\ \colrule
~~~~ 0.10$\pm$0.05  &  -----  & 0.13$\pm$0.07 & [\cite{Braun:1986ty}]  \\
~~~~   -----        &  0.06$\pm$0.03 & -----  &   [\cite{Balitsky:1989jb}]  \\  ~~~~ 0.16$\pm$0.08.   &  0.22$\pm$0.12 & 0.16$\pm$0.08 & [\cite{Ross:1993gb}] \\ ~~~~   -----        &  0.03$\pm$0.01 & -----  & [\cite{Stein:1994zk}] \\
~~~~   -----        &  0.03$\pm$0.06 & -----  & [\cite{Ioffe}] \\
~~~~   0.08$\pm$0.04 &      0.09$\pm$0.03 & -----  & [\cite{Balla:1997hf}] \\
~~~~   -----        &  -----  & 0.16$\pm$ 0.08  &  [\cite{Weiss:2002qq}]  \\
~~~~   -----        &  0.10$\pm$0.07  & -----   &  [\cite{SW}]    \\ 
\botrule
\end{tabular}}
\end{table}
The numbers from  Refs. \refcite{Braun:1986ty}-
\refcite{Ioffe} were obtained  using three-point function QCD sum rules method
with different interpolating currents. The results \cite{Balla:1997hf},  
\cite{Weiss:2002qq} were obtained using an instanton model of 
the non-perturbative vacuum. 
Within the quoted  error bars they all  are consistent.
However, the work of Ref. \refcite{Ioffe}  
demonstrates the importance of careful estimates of theoretical 
uncertainties and is  putting a  huge question  mark next to the small 
result of Ref.\cite{Stein:1994zk}, which is significantly  smaller 
than the one  from Ref.\refcite{Balitsky:1989jb}. 
Moreover, phenomenological determination of the parameter  ${\rm B}$ 
from the polarized structure function data by means of integrating 
the  $h(x)/Q^2$ model 
extracted from the data  \cite{SW} 
supports  the result  of the  QCD sum rules 
analysis of Ref.\refcite{Balitsky:1989jb}.

Thus the numbers presented in Table 1 and discussions below it  
lend  support 
to 
Eq. (\ref{nonpert}) and together with the perurbative equation of 
Ref.({\ref{pertuv}) 
are consistent with the main relation of   Eq.(1)
discovered in Ref.\refcite{Kataev:2005ci} and 
discussed in detail in  
Ref.\refcite{Kataev:2005hv}. 
However, more careful checks are  still needed.
One of them is related to
the necessity of  independent calculations of 
the  parameters ${\rm A}$, ${\rm B}$ and ${\rm C}$.
Another one is presumes  more detailed studies of the 
outcomes  of taking Borel transform  from the Borel 
images, as calculated in Refs.\refcite{Broadhurst:1993ru},
\refcite{Broadhurst:2002bi}. The first results, presented in 
\cite{Brooks:2006it} for $N_f=0$ seem to  confirm  Eq.(1) in the region  
 $Q^2/\Lambda^2\geq 6$ which approximately corresponds 
to the region of energies considered in 
\cite{Kataev:2005ci},\cite{Kataev:2005hv}.

Consider now  the experimental consequences 
of Eq.(1). We will use experimental values of 
CCFR-NuTeV collaboration for the ${\rm{Q^2}}$-dependence of the 
$\rm{GLS}$ sum rule \cite{Kim:1998ki}, extract from them the 
 ${\rm Q^2}$ dependence of the  ${\rm Bp}$ sum rule using the 
approximate theoretical relation of Eq.(1) and compare these results with the 
concrete experimental data (see Table 2).
\begin{table}
\tbl{The comparison of ${\rm Bp}$ results extracted from   
 ${\rm GLS}$ values  using  Eq. (1) with the 
direct experimental extractions of  ${\rm Bp}$}
{\begin{tabular}{@{}lcr@{}}
\toprule
~~~~~~$\rm{Q^2}$   &   $\rm{ Bp}$ from Eq.(1)  
& $\rm{Bp}$  ~(expt)   
  \\ 
\colrule
~~~~ 2.00  &  0.174$\pm$ 0.012 & 0.169$\pm$0.025~~~[\cite{Abe:1998wq}]  \\
~~~~ 3.16  &  0.178$\pm$0.008  & 0.164 $\pm$0.023~~[\cite{Abe:1998wq}] \\    
~~~~ 5.01  &  0.195$\pm$0.014  & 0.181$\pm$0.022~~~[\cite{Adeva:1998vw}] \\ 
~~~~ 12.5  &  0.196$\pm$ 0.016 & 0.195$\pm$0.029~~~[\cite{Adeva:1997is}]  \\
\botrule
\end{tabular}}
\end{table}
One can see that within existing error bars 
the results for 
${\rm Bp}$ sum rule 
motivated by 
the ${\rm GLS}$ sum rule 
experimental numbers the ones  
based on the exact experimental measurements 
are in 
good agreement. 
Moreover  extracted by theoreticians from existing SLAC and SMC data,
$\rm{ Bp(3~GeV^2)}$=0.177$\pm$0.018 \cite{Altarelli:1996nm}
which, within error bars, does  not contradict  the value 
$\rm{ Bp(3~GeV^2)}$=0.164$\pm$0.011  \cite{Ellis:1995jv},
and 
is in beautiful ( although presumably  accidental) agreement  with 
the result of applying Eq.(1) to the experimental value 
of the ${\rm GLS}$ sum rule (see second entry in Table 2).

In order to estimate better the 
errors and limitations of 
 Eq.(1), it is highly desirable to try to extract the  
$Q^2$ dependence  
of the ${\rm GLS}$ sum rule from  already existing NuTeV 
data for $xF_3$.  It would also be  interesting to get new data for the 
${\rm Bp}$ sum rule.
These data may be obtained  at JLAB and 
by COMPASS Collaboration at CERN, if it  will 
be able to continue running this  experiment  using hydrogen target.
Another interesting application of   Eq.(1) would be  the 
estimation of  the 
${\rm Q^2}$ dependence of the  still experimentally unmeasured  ${\rm Bup}$   
sum rule. This proposal was made whilst  planning for the 
 hadronic program of Neutrino Factories 
\cite{Mangano:2001mj},\cite{Alekhin:2002pj}.

I am grateful to I.A. Savin 
for discussions. It is the pleasure to thank D.J.Broadhurst 
for fruitful collaboration and  
the members and visitors 
to IPPP, Durham, UK,
where this write-up of my  talk at ICHEP-06 was prepared,   
for the interest in its content.  The work is supported by  
RFBR Grants 05-01-00992 and 06-02-16659.

\end{document}